\documentclass[preprint, aps, amsmath]{revtex4}
\usepackage[dvips]{graphics}

\newcommand{\be}{\begin{equation}}
\newcommand{\ee}{\end{equation}}

\newcommand{\bea}{\begin{eqnarray}}
\newcommand{\eea}{\end{eqnarray}}
\newcommand{\bd}{\begin{displaymath}}
\newcommand{\ed}{\end{displaymath}}
\newcommand{\bi}{\begin{itemize}}
\newcommand{\ei}{\end{itemize}}
\newcommand{\bc}{\begin{center}}
\newcommand{\ec}{\end{center}}
\newcommand{\bfl}{\begin{flushleft}}
\newcommand{\efl}{\end{flushleft}}
\newcommand{\bfr}{\begin{flushright}}
\newcommand{\efr}{\end{flushright}}
\newcommand{\f}{\frac}

\def\ra{\rightarrow}
\def\6{\partial}

\def\={\!\!\!&=&\!\!\!}
\def\+{\!\!\!&&\!\!\!+~}
\def\-{\!\!\!&&\!\!\!-~}

\begin{document}
\title{Universality of the conductance  in quantum dot transport}
\author{M. Crisan}
\affiliation{Department of Theoretical Physics, ``Babe\c{s}-Bolyai" University, 400084 Cluj-Napoca, Romania}
\author{I. Grosu}
\affiliation{Department of Theoretical Physics, ``Babe\c{s}-Bolyai" University, 400084 Cluj-Napoca, Romania}
\author{I. \c{T}ifrea}
\affiliation {Department of Physics and Astronomy,
California State University Fullerton, Fullerton, CA 92384, USA }

\begin{abstract}
We revisited the scaling behavior of the transport properties of a quantum dot system described by the spin-1/2 Anderson model using analytical methods. In the  low temperature limit  we show that the conductance has a universal behavior with universality between temperature and bias. We compare this result with the empirical formula used to fit the experimental data for conductance in the case of the equilibrium transport through a single channel quantum dot. In the high temperature limit the conductance obtained from the Anderson model is compared with previous results obtained from the Kondo model. The universal behavior is present also in the high temperature limit. These results are in good agreement with the Renormalization group calculations.
\end{abstract}

\maketitle

\section{Introduction}
Recent experimental results demonstrated the importance of the electronic correlations for the behavior of the low temperature transport of quantum dots and break junctions devices containing $C_{60}$. A quantum dot (QD) consist of a confined region of electrons of size $l \sim \lambda_{B}$, ($\lambda_{B}$ - the de Broglie  wave-length), coupled to leads via tunnel barriers. The conduction of such a device was described by the well known Anderson model \cite{an,he1}, in terms of the interacting constants $\Delta= \pi N(0)|V_{kd}|^{2}$ ( $V_{kd}$ - the interaction between the conduction electrons and $d$-impurities) and Coulomb interaction $U$. Most of the transport studies for such a configuration are carried in the $T=0$ limit, in the case of strong ($U/\Delta \gg 1$) and weak ($U/\Delta\ll 1$) coupling between the conduction and localized electrons. In the strong coupling and zero temperature limit the Coulomb interaction $U$ provides the largest energy scale. Although the $T=0$ case provides a reasonably answer for the transport problem, the case of a finite temperature ($T\neq 0$) should be carefully considered as the transport properties of the system can be significantly influenced by thermal effects. For $\Delta<T<U$ quantum fluctuations are small and transport  is dominated by charge effects. This regime  is well described by the concept known as Coulomb blockade and in this situation the temperature dependence of the conductance $G(T)$ was calculated by Beenaker \cite{be}.  At low temperature  the conductance presents a series of equidistant peaks as function of gate voltage $V$ separated by a spacing $U$. These peaks correspond to a fractional number of electrons in the dot and separate the Coulomb blockade valleys corresponding to either an even or odd number of electrons. In the regime $T<< \Delta<< U$ the strong quantum fluctuations can lead to important  modifications  of the Coulomb blockade scenario. For an odd number  of electrons in the quantum dot (called now a Kondo quantum dot) the system can have a net spin $1/2$ and the Kondo effect can develop. As a consequence the conductance in the odd electron valleys is enhanced, turning those valleys into plateaus of near perfect transmission.

The experimental measurement of the system's conductance at temperatures below the Kondo temperature ($T_K$) have been a difficult task for physicists. In particular, the low temperature regime defined above ($T\ll\Delta$) requires a direct control over the interaction between localized and conduction electrons. Recent advances in nanotechnologies allow a better control over the geometry of the system's electrodes and implicitly over the electron interactions. The temperature dependence of the system's conductance \cite{go,yu,pa} have been fitted using the empirical formula
$G_{EK}(T)=G_{0}[1+2^{1/s-1}(T/T^{'}_{K})^{2}]^{-s}$ where  $T^{'}_{K}=T_{K}/(2^{1/2s}-1)^{1/2}$ with $s=0.22$ and
$G_{EK}(T_{K})=G_{0}/2$. On the other hand, in the low temperature limit, theoretical estimations of the system's conductance lead to $G(T)=G_{0}[1-\pi^{2}(T/T_{K})^{2}/16]$ with $G_{0}=2e^{2}/h$; this result is given by the temperature dependence of the Kondo resonant peak in the system's density of states. Although there is a good agreement between experimental and theoretical results in the very low temperature limit ($T\ll T_K$), in the limit of higher temperatures experimental and theoretical data  behave differently.

Here, we revisit the problem of universal behavior of the transport properties in quantum dot systems. In the equilibrium case the scaling behavior was obtained using the microscopic Anderson model with a finite on-site Coulomb interaction $U$ in the non-crossing approximation \cite{ge}, but this method can give only qualitative results. On the other hand, the origin of the differences between the theoretical conductance $G(T)$ obtained in the Fermi-liquid approximation and the experimental conductance $G_{EK}$ was not discussed in the literature. In Section II we present the calculation of the system's conductance in the equilibrium case for finite on-site Coulomb interaction $U$ and discuss the possible origin of the enhancement in the Kondo temperature. The universal behavior for the non-equilibrium case is presented in Section III using the microscopic Anderson model. This problem was treated first by Appelbaum  \cite{app} and Anderson \cite{an2} and more recent  reconsidered by Schiller and Hersfield \cite{sch} using the Touluse exact model. The same situation, but for the case of low and high bias voltages, was reconsidered by Oguri \cite {og1,og2} using the Yamada $U$-expansion \cite{ya}. Using a similar model we reconsider the problem in connection with recent experimental results obtained by Grobis et al. \cite{gr} and recalculate the scaling function
of the system's conductance. In Section IV we discuss our results in connection with other results from the literature.

\section{Universal scaling in  equilibrium  transport through a quantum dot}
In this section we will present a detailed analysis of the universal behavior of a single channel quantum dot using the Anderson model. We will consider both low temperature ($T\ll T_K$) and high temperature ($T\gg T_K$) regimes. We will discuss the possible origin of the differences between the theoretical and empirical fits to the system's conductance in the low temperature regime.

\subsection{Model and Green functions}
We start from the simple symmetric Anderson model described by the Hamiltonian:

\begin{equation}
H=H_{lead}+H_{dot}+H_{int}
\end{equation}
where:
\begin{equation}
H_{lead}=\sum_{k\sigma}\epsilon_{k}c^{\dag}_{k\sigma}c_{k\sigma}\;,
\end{equation}
\begin{equation}
H_{dot}= \sum_{\sigma}\epsilon_{d}d^{\dag}_{\sigma}d_{\sigma}
+U\left(d^{\dag}_{\uparrow}d_{\downarrow}-\f{1}{2}\right)\left(d^{\dag}_{\uparrow}d_{\downarrow}-\f{1}{2}\right)
-\frac{U}{4}\;,
\end{equation}
and
\begin{equation}
H_{int}= \sum_{k\sigma}V_{kd}(c^{\dag}_{k\sigma}d_{\sigma}+h.c)\;.
\end{equation}
This Hamiltonian has been used by Anderson to describe the occurrence of the magnetic moments in dilute alloys. The first part of the Hamiltonian, $H_{lead}$, describes the conduction electrons in the leads in terms of the creation ($c^\dagger_{k\sigma}$) and annihilation ($c_{k\sigma}$) operators. The second part of the Hamiltonian, $H_{dot}$, similar to the Anderson's model $d$-localized level, describes the electrons localized in the quantum dot in terms of the operators $d^\dagger_\sigma$ and $d_\sigma$. Finally, $H_{int}$ describes the interaction between the conduction electrons and the localized electrons.

In order to calculate the physical quantities described by this Hamiltonian we will use the Green function:
\begin{equation}
G^{-1}(i\omega_{n})=G_{0}^{-1}(i\omega_{n})-\Sigma(i\omega_{n})\;,
\end{equation}
where $\omega_{n}=(2n+1)\pi T$. $G_{0}(i\omega_{n})$, the bare Green's function in the absence of the Coulomb on-site interaction $U$, can be calculated as:
\begin{equation}
G_{0}(i\omega_{n})=\frac{1}{i\omega_{n}-\epsilon_{d}+i\Delta sgn\omega_{n}}\;,
\end{equation}
where $\Delta=\pi N(0)|V_{kd}|^{2}$ with $N(0)$ being the density of states at the Fermi level for the conduction electrons. The self-energy $\Sigma(i\omega_n)$ describes the effect of the on-site Coulomb interaction and can be evaluated using the Yamada perturbative $U$-expansion \cite{ya} to the ($\omega^2$, $T^2$) order in the region $U/(\pi\Delta)\simeq 1$ and $0<\epsilon_d/\Delta<1$. The energy scale of the model, for the $U=0$ case, is given by the Kondo temperature $T_{K}$:
\begin{equation}
T_{K}=\Delta \exp{\left(\frac{\pi E_{d}}{\Delta}\right)}\;,
\end{equation}
where $E_{d}=\epsilon_{d}+\Delta\ln (D/\Delta)/\pi$,  and $D$ is the bandwidth. Note that $T_{K}$ given by this expression has to be considered only as an energy scale, a better expression for the Kondo temperature, which can be used to fit the experimental result, being $T^{r}_{K}=\sqrt{\Delta T_{K}}$.

The  density of states  corresponding to the localized electrons in the quantum dot can be obtained from their Green function as:
\begin{equation}
\rho(\omega, T)=-\frac{1}{\pi} \; Im G(\omega, T)=\frac{1}{\pi}\;
\frac{Im \Sigma(\omega,T)} {\left[\omega-E_{d}(\omega,T)\right]^{2}+
\left[Im\Sigma(\omega, T)\right]^{2}}\;,
\end{equation}
where $E_{d}(\omega,T)=\epsilon_{d}+Re\Sigma(\omega, T)$. In the following  we will consider two different situations,  the case of low temperatures, $(\omega,T)<<T_{K}$, and the case of high temperatures, $(\omega, T)>>T_{K}$.

\subsubsection{Low temperature regime}

In this case $Re\Sigma(\omega, T)\simeq -\epsilon_{d}$ and $Im \Sigma(\omega,T)=\Delta+ \Sigma_{1}(\omega,T)$ where $\Sigma_{1}(\omega)$ is given by:
\begin{equation}\label{sigma1}
\Sigma_{1}(\omega,T)=a\Delta \; \frac{\omega^{2}+(\pi T)^{2}}{T^{2}_{K}}
\end{equation}
The value of the constant $a=\pi^{4}/16$ was given by Costi \textit{et. al.} \cite{co}, for the symmetric Anderson model using the Yamada \cite{ya} expansion of the self-energy and is exactly the value obtained in the Fermi liquid approximation by Nozi\`{e}res \cite{no}. Using this value for the constat $a$, the resulting temperature dependence of the system's conductance describes the experimental results only qualitatively. An analytical expression of the constant $a$ from the self-energy, which can change quantitatively the Kondo temperature, was given by Kirchner \textit{et al.} \cite{ki}:
\begin{equation}
 a=\frac{\pi^{4}W^{2}}{8e^{(3/2+C)}}\; \frac{R-1}{\sin^{2}(\pi n_{d\sigma})}\; \frac{|\epsilon_{d}|}{D}\;,
\end{equation}
where $C=0.57721$ is the Euler constant, $W=0.4128$ is the Wilson number, $R=2$ is the Wilson ratio,  and $2D$ is the bandwidth of the conduction electrons. The important parameter of the model is the product $a\Delta$ which for the Fermi liquid model was calculated as $a\Delta=0.0239\;D$. Using the result given in Ref. \cite{ki},  taking for $|\epsilon_{d}|/D=0.81$ and $\Delta/D=0.2$ we obtained $a\Delta=0.0236$, which is  a reasonable value compared with the value from the Fermi liquid model. We can get information about the magnitude of this parameter from the experimental data obtained for the  temperature dependence of the Kondo resonance width \cite{na}. Based on Eq. (\ref{sigma1}) the imaginary part of the self-energy can be written as:
\begin{equation}
Im\Sigma(T,\omega)=\Delta \left[1+\left(\frac{\omega}{T^{\ast}_{K}}\right)^{2}+\left(\frac{\pi T}{T^{\ast}_{K}}\right)^{2}\right]\;,
\end{equation}
where $T^{\ast}_{K}=T_{K}/\sqrt{a}$. Using the above expression the electronic density of states corresponding to the QD electrons can be approximated as:
\begin{equation}
\rho(T, \omega)\simeq\left[\pi\Delta\left(1+\left(\frac{\omega}{T^{\ast}_{K}}\right)^{2}+\left(\frac{\pi
T}{T^{\ast}_{K}}\right)^{2}\right) \right]^{-1}\;,
\end{equation}
or if we introduces the notation $\gamma^{2}(T, T_{K}^{\ast})=(\pi T)^{2}+(T^{\ast}_{K})^{2}$, as:
\begin{equation}
\rho(T, \omega)=\frac{1}{\pi\Delta}\left[1+\left(\frac{\pi
T}{T^{\ast}_{K}} \right)^{2}\right]^{-1}\frac{\gamma^{2}(T, T^{\ast}_{K})}{\gamma^{2}(T, T^{\ast}_{K})+\omega^{2}}\;.
\end{equation}
This form for the density of states is specific for the Fermi liquid behavior and describes  the low temperature domain defined as  ${(\omega,T)<<T_{K}}$.

\subsubsection{High temperature regime}
In the high temperature regime, $\omega\gg T_K$, the effects of the Kondo state on the system's conductance can be evaluated using the electronic self-energy in terms of the t-matrix approximation, $\Sigma(\omega)\simeq t(\omega)$. The electronic t-matrix can be calculated using the Anderson model along with an appropriate decoupling in the equation of motion, a method which is equivalent to the calculation of the t-matrix with respect to a perturbation expansion up to the sixth order in the interaction term $V_{kd}$ \cite{ma}. The connection between the electronic $t$-matrix and the $d$-electrons density of states was discussed first by Langreth \cite{la} and is given by
\begin{equation}
Im \; t(\omega)=-\frac{\Delta}{N(0)}\rho(\omega,T)\;.
\end{equation}
In the high temperature limit, for a correct description of the system's properties, the $t$-matrix approximation has to include effects related to the inelastic scattering of the conduction electrons on the spins attached to the $d$-electrons in the quantum dot. Following Mamada and Shibata \cite{ma} the evaluation of the $t$-matrix for the symmetric Anderson model can be done in the $U\ra\infty$ limit:
\begin{equation}
Im \; t(\omega)=- \pi
N(0)\left[\frac{3}{16}J^{2}(\omega)-\frac{3}{8}N(0)J^{3}(\omega)\ln\left(\frac{D}{|\omega|}\right)\right]\;'
\end{equation}
where the running coupling constant  $J(\omega)=1/\left(\pi N(0)\ln\frac{|\omega|}{T_{K}}\right)$ in the high temperature ($|\omega|>>T_{K}$) regime. Based on the above equations the electronic density of states can be approximated as:
\begin{equation}
\rho(\omega,T)\simeq
\frac{1}{\Delta}\;\frac{3}{16}\;\frac{1}{\ln^{2}\frac{\omega}{T_{K}}}\;.
\end{equation}

\subsection{Conductance}

The conductance is one of the most investigated physical quantities in quantum dot systems. For the transport properties of a single quantum dot system we expect a universal behavior of the conductance $G(T/T_{K})$, behavior which is specific to the Kondo problem. However, the experimental facts prove that there are deviations from the specific Fermi liquid behavior  in the low temperature regime, $T\ll T_K$. As a consequence the experimental data were fitted with an empirical formula $G_{EK}(T)$. In the following we will calculate the conductance behavior for this regime using the Anderson model. On the other hand the high temperature behavior is obtained taking the results from the perturbation theory applied to the Anderson model.

\subsubsection{Low temperature regime}
To obtain the temperature dependence of the system's conductivity in the low temperature regime we start with the general equation for the current:
\begin{equation}
I(V, T)=\frac{\pi e \Delta}{h}\int_{-\infty}^{\infty}d\omega\rho(\omega)[f(\omega-eV)-f(\omega+eV)]\;,
\end{equation}
where $V$ is the applied external bias and $f(\omega)$ is the Fermi function. The integral in the above definition can be calculated and using the definition of the system's conductance,  $G(V,T)=dI(V,T)/dV $, we find:
\begin{equation}
G(V, T)=\frac{G_{0}}{2\pi}\frac{(T^{\ast}_{K})^{2}}{\gamma T}Re\left[\Psi^{(1)}\left(\frac{1}{2}+\frac{\gamma+ieV}{2\pi T}\right)\right]\;,
\end{equation}
where $\gamma\equiv \gamma(T, T^{\ast}_{K})$ and $\Psi^{(1)}(z)=d\Psi(z)/dz$ is the trigamma function. Using the asymptotic behavior of the trigamma function, $\Psi^{(1)}(z)\simeq 1/z +1/2z^{2}$, in the low voltage limit $eV/{k_BT_K}\ll 1$, the system's conductance can be approximated as
\begin{equation}
G(0,T)\simeq G_{0}\left[1-2\left(\frac{\pi T}{T^{\ast}_{K}}\right)^{2}\right]\;,
\end{equation}
a result which is in good agreement with Fermi-liquid behavior, i.e., the conductance decreases quadratically
with temperature $T$ in the low temperature regime, $T<<T^{\ast}_{K}$. This expression for the system's conductance is very similar to the empirical formula used in fitting the experimental data $G_{EK}(0, T)=G_{0}\left[1-c_{T}\left(T/T_{K}\right)^{2}\right]$, where $c_{T}$ is determined by the definition of the Kondo
temperature $G_{EK}(T_{K})=G_{0}/2$.

\subsubsection{High temperature regime}

In the high temperature regime, $\omega>>T_{K}$, the system's conductance $G(T)$ can be evaluated using the general relation \cite{af}:
\begin{equation}
G(T)=G_{0}\int_{-\infty}^{\infty}d\omega\left[-\frac{df(T)}{d\omega}\right][-N(0)Im t(\omega)]\;,
\end{equation}
a relation which combined with the expression for the electronic $t$-matrix leads to:
\begin{equation}
G(T)\simeq G_{0}\frac{3\pi^{2}}{16}\frac{1}{\ln^{2}\frac{T}{T_{K}}}\;,
\end{equation}
where the $J^{3}(\omega)$-term has been neglected. This result was obtained  first by Kaminski \textit{et al.} \cite{ka} from the Kondo Hamiltonian.  We mention that in the recent analysis by Borda \textit{et al.} \cite{bo} the inelastic scattering gives a $1/\ln^{2}(\omega/T_{K})$ term with the  increasing of the parameter $U/\Delta$.  This conclusion is in agreement with the  above calculation of the conductance since the Kondo model is  equivalent to the Anderson model in the limit of large $U/\Delta$  and $\omega<< U$.

\section{Universal scaling in non-equilibrium transport through a quantum dot}

In this section we analyze the nonequilibrium transport properties of a quantum dot system. We will consider in detail the low temperature regime, $T\ll T_K$, using the same formalism as in the equilibrium case to identify the possible universal behavior of the system's conductance, occupancy of the quantum dot, and magnetic spin susceptibility in the nonequilibrium case.

\subsection{Conductance}
Following the same procedure as in the equilibrium case, we will need to evaluate first the electronic density of states for our system. The electron Green's function for the nonequilibrium case was calculated in the electron-hole symmetric limit using the Hewson-Renormalization theory \cite{he} by Oguri as \cite{og1,og2}:
\begin{equation}
\overline{G^{r}}(\omega)=\frac{z}{\omega+i\overline{\Delta}+i\frac{
\overline{U}^{2}}{2\overline{\Delta}(\pi\overline{\Delta})^{2}}[\omega^{2}+\frac{3}{4}(eV)^{2}+
(\pi T)^{2}]}\;,
\end{equation}
where $\overline{\Delta}=z\Delta$. The parameters in the above equation are given by $z=1-(3-\pi^{2}/4)u^{2}$, $u=U/(\pi\Delta)$,  and $\overline{U}=z^{2}\Gamma_{\uparrow,\downarrow}(0,0)$, $\Gamma_{\sigma,\sigma'}(\omega,\omega)$ being the renormalized vertex function. Accordingly, the system's density of states defined as $\rho(\omega, T)=-Im \overline{G^{r}}(\omega)/\pi$ can be approximated as:
\begin{equation}
\rho(\omega,T)\simeq\frac{1}{\pi\overline{\Delta}}\;\frac{1}{1+\frac{1}{2\overline{\Delta}^{2}}
(\frac{\overline{U}}{\pi\overline{\Delta}})^{2}[\omega^{2}+\frac{3}{4}(eV)^{2}+(\pi T)^{2}]}\;.
\end{equation}

Following the same procedure as in the equilibrium case, in order to evaluate the system's conductance we will start from the estimation of the nonequilibrium current in the presence of an external voltage. Starting from the general definition we find
\begin{equation}
I(V)=\frac{e}{2}\int_{-\infty}^{\infty}d\omega\frac{\tanh(\frac{\omega-eV}{2T})-\tanh(\frac{\omega-eV}{2T})}
{1+a[\omega^{2}+\frac{3}{4}(eV)^{2}+(\omega T)^{2} ]}\;,
\end{equation}
where $a=1/(2\overline{\Gamma}^{2})(\overline{U}/\pi\overline{\Gamma})^{2}$. This integral can be performed exactly and we obtain :
\begin{equation}
I(V)=\frac{2e}{a\alpha}Im\Psi\left[\frac{1}{2}+\frac{\alpha+ieV}{2\pi T}\right]\;,
\end{equation}
where $\Psi(x)$ is the digamma function and $\alpha^{2}=1+\frac{1}{a}+\frac{3}{4}(eV)^{2}+(\pi T)^{2}$. The imaginary part of the di-gamma function can be evaluated and the current $I(V)$ becomes:
\begin{equation}
I(V)=\frac{2e}{h}\frac{\sqrt{2}T_{K}}{\left[1+\frac{3}{8}\left(\frac{eV}{T_{K}}\right)^{2}+\frac{\pi^{2}}{2}
\left(\frac{T}{T_{K}}\right)^{2}\right]}\arctan\frac{(\frac{eV}{\sqrt{2}T_{K}})}
{\left[1+\frac{3}{8}\left(\frac{eV}{T_{K}}\right)^{2}
+\frac{\pi^{2}}{2}\left(\frac{T}{T_{K}}\right)\right]^{1/2}}\;,
\end{equation}
or in the low voltage limit ($V\ra 0$)
\begin{equation}
I(V)\simeq\frac{2e^{2}}{h}V\frac{1}{1+\frac{3}{8}(\frac{eV}{T_{K}})^{2}+\frac{\pi^{2}}{2}(\frac{T}{T_{K}})^{2}}\;.
\end{equation}
Accordingly, the system's conductance exhibits a scaling behavior of the form $[G(T,0)-G(T,V)]/G_0=F(eV/T_K, T/T_K)$ with
\begin{equation}
F\left(\f{eV}{T_K},\f{T}{T_K}\right)=\frac{3}{8}\left(\frac{eV}{T_{K}}\right)^{2}-
\frac{3\pi^{2}}{16}\left(\frac{T}{T_{K}}\right)^{2}\left(\frac{eV}{T_{K}}\right)^{2}\;,
\end{equation}
a result  which is qualitative agreement with Ref. \cite{gr}.

\subsection{Particles number and magnetic susceptibility}
Additional parameters show an universal behavior in quantum dot systems. For example, for the single dot system the occupancy (number of particles) of the main quantum dot and the magnetic susceptibility of the system can be shown to present an universal behavior. Let consider for the beginning the occupancy of the main dot of the system. By definition
\begin{equation}
n_{d}=-\frac{2}{\pi}\int_{-\infty}^{\infty}d\omega f(\omega)Im \overline{G^{r}_{d}(\omega)}\;,
\end{equation}
where $f(\omega)$  represents the Fermi function and $\overline{G^{r}_{d}(\omega)}$ is the electronic Green's function. The integral on the right hand side of the above equation can be performed and we get:
\begin{equation}
n_{d}= \frac{2}{(R-1)^{2}\sqrt{\frac{2}{(R-1)^{2}}+\frac{3}{4}(\frac{eV}{\overline{\Delta}})^{2}}}\;,
\end{equation}
where we used the Wilson ratio $R$ defined as $R=1+\overline{U}/(\pi\overline{\Delta})$. The system's magnetic susceptibility $\chi(T)$ can be calculated as:
\begin{equation}
\chi(T)=\frac{(g\mu_{B})^{2}}{2T}\int_{-\infty}^{\infty}d\omega\rho(\omega,T)f(\omega)[1-f(\omega)]\;.
\end{equation}
The calculation is relatively simple and we obtain
\begin{equation}
\chi(T,V)= \frac{(g\mu_{B})^{2}}{\pi\overline{\Delta}}\frac{1}{1+\frac{(R-1^{2})}{2}\left[\frac{3}{4}
\left(\frac{eV}{\overline{\Delta}}\right)^{2}+\pi^{2}\left(\frac{T}{\overline{\Delta}}\right)^{2}\right]}\;.
\end{equation}
Note that for $\overline{\Delta}=T_{K}$ and $R=2$ we obtain the scaling behavior in the Kondo regime.

\section{Summary and Conclusions}

We  studied the universal behavior of a single quantum dot system at low temperatures described by the microscopic Anderson model using the many body methods. The main results can be summarized as follows:
\begin{itemize}
\item the temperature dependence of the system's conductance in the equilibrium case is given  by the second order corrections in the Coulomb interaction by an effective temperature dependent density of states. In the low temperature approximation,  $T<<T^{\ast}_{K}$, the correction has the form $G(T/T^{\ast}_{K})\sim a(T/T^{\ast}_{K})^{2}$, where the constant $a$ depends on the model's parameters. This result can be a reasonable explanation for the existence of quantitative differences between the experimental data and the formula obtained using the standard Fermi liquid. The empirical formula $G_{KE}(T)$ does not contains a new physical picture, but new numerical coefficients which may be given by a more realistic model.

\item in the high temperatures limit, $T >>T^{\ast}_{K}$, the system's transport properties in the equilibrium case were obtained using the $t$-matrix approximation for the Anderson model. The system's conductance was calculated as $G(T/T^{\ast}_{K})\sim  1/\ln^{2}(T/T^{\ast}_{K})$, a result obtained previously in the literature using the Kondo Hamiltonian \cite{ka}. This result shows that in the higher order of the perturbation theory we can obtain, using a microscopic Hamiltonian, the same result as from the phenomenological Kondo Hamiltonian.

\item the scaling behavior of the system's conductance in the equilibrium case was analyzed in literature using the Hewson-Renormalization procedure for the Anderson model. In our work, we obtained results similar to the proposed phenomenological description of the universal scaling in non-equilibrium transport (Grobis \textit{et al.}  \cite{gr}) using the microscopic model and many-body methods. These results reduces to the equilibrium case for the zero bias case.
\end{itemize}

\end{document}